\documentclass{article}

\usepackage{epsfig}

\title{Chiral Resonant Solitons in
 Broer-Kaup Type New Hydrodynamic Systems}

\author{Jyh-Hao Lee and Oktay K. Pashaev\\Institute of Mathematics, Academia Sinica
   Taipei, Taiwan\\ (e-mail: leejh@ math.sinica.edu.tw)
 \\ Department of Mathematics, Izmir Institute of Technology \\ Urla-Izmir, 35430, Turkey (e-mail: oktaypashaev@iyte.edu.tr)}

\begin{document}
\maketitle

\begin{abstract}                
New Broer-Kaup type systems of hydrodynamic equations are derived from the
derivative reaction-diffusion systems arising in SL(2,R) Kaup-Newell hierarchy, represented
in the non-Madelung hydrodynamic form. A relation with the problem of chiral solitons in quantum potential as a
dimensional reduction of 2+1 dimensional Chern-Simons theory for anyons is shown.
By the Hirota bilinear method, soliton solutions
are constructed and the resonant character of soliton interaction is found.

\end{abstract}


\section{Introduction}
Recently,  a modification of the nonlinear Schr\"odinger (NLS) equation by a quantum potential has been studied in several
problems arising in low dimensional gravity, \cite{Pashaev et al.a}, plasma physics, \cite{Lee et al.a}, the capillary wave, and information theory, \cite{Parwani et al.}. Subsequently, the influence of this potential
on anyons in 2+1 dimensions has been studied \cite{Pashaev et al.c}, and
the Abelian Chern-Simons gauge field interacting with NLS has been  represented as a planar Madelung fluid \cite{Lee et al.c} , where the Chern-Simons Gauss law has the simple physical meaning of creation of the local vorticity for the flow. For the static flow when the velocity of the center-of-mass motion is equal to the quantum velocity, the fluid admits an N-vortex solution. It turns out that in this theory the Chern-Simons coupling constant and the quantum potential strength  are quantized.

 Reduction of problem to 1+1 dimensions leads to JNLS and some versions of Derivative NLS in quantum potential.
Hence the chiral solitons appear as solutions of the derivative NLS with quantum potential \cite{Lee et al.b} .  The last one
by the Madelung transform is represented as the derivative Reaction-Diffusion (DRD) system, arising in SL(2,R) Kaup-Newell hierarchy, and giving rise to the resonant soliton phenomena
\cite{Pashaev et al.d}..

In the present paper by using new, the non-Madelung representation,
we formulate the problem in terms of novel hydrodynamic systems of the Broer-Kaup type. Then by Hirota's bilinear method we
construct chiral solitons for the system and show the resonance character of their interaction.

\section{Dimensional reduction of Chern-Simons theory }

We consider the Chern-Simons gauged Nonlinear Schr\"odinger model (the Jackiw-Pi model) with nonlinear quantum potential term of strength $s$ \cite{Pashaev et al.c}:

\begin{equation} L = {\kappa \over 2}\epsilon^{\mu\nu\lambda}A_{\mu} \partial_{\nu}A_{\lambda}
 + {i \over 2}(\bar\psi D_{0}\psi - \psi \bar D_{0}\bar \psi)
- \bar{\bf D}\bar\psi {\bf D}\psi + s\nabla |\psi| \nabla |\psi| + V(\bar\psi \psi), \label{CS}\end{equation}

 where $D_{\mu} = \partial_{\mu} + ieA_{\mu}$,
$(\mu = 0,1,2)$.
Classical equations of motion are

\begin{equation}iD_{0}\psi + {\bf D}^{2}\psi + V'\psi = s \frac{\Delta |\psi|}{|\psi|}\psi, \label{CSE1}\end{equation}
\begin{equation}\partial_{1}A_{2} - \partial_{2}A_{1} = {e \over \kappa}\bar\psi\psi, \label{CSE2}\end{equation}

\begin{equation}\partial_{0}A_{j} - \partial_{j}A_{0} = -{e \over \kappa}i\epsilon_{jk} (\bar\psi D_{k}\psi - \psi \bar D_{k}\bar\psi),\,\, (j,k = 1,2). \label{CSE3}\end{equation}

We consider dimensional reduction of this model when all field are independent of $x_2$ space variable, so that $\partial_2 = 0$.  Then in terms of $\tilde A_0 \equiv A_0 + e A_2^2$, and $B \equiv A_2$, we obtain

\begin{equation}i(\partial_0 + ie  A_0)\psi + (\partial_1 + ie A_1)^{2}\psi + V'\psi = s \frac{\partial^2_1|\psi|}{|\psi|}\psi, \label{CSRE1}\end{equation}

\begin{equation} \partial_0 A_1 - \partial_1  A_0 = 0 \label{CSRE2}\end{equation}

\begin{equation} \partial_1 B = \frac{e}{\kappa} \bar\psi \psi \label{CSRE3}\end{equation}

\begin{equation} \partial_0 B = i\frac{e}{\kappa} \left[\bar\psi (\partial_1 + ieA_1) \psi - \psi (\partial_1 - ie A_1) \bar\psi\right] \label{CSRE4}\end{equation}

Here and below we skip the tilde sign for $A_0$.
The last two equations are compatible due to (\ref{CSRE1}) and the corresponding continuity equation

\begin{equation} \partial_0(\bar\psi \psi ) = i \partial_1 \left[\bar\psi (\partial_1 + ieA_1) \psi - \psi (\partial_1 - ie A_1) \bar\psi\right]\label{cont} \end{equation}

implies compatibility of equations (\ref{CSRE3}),(\ref{CSRE4}).
Integrating these equations we find $B$ in terms of density $\bar\psi \psi$

\begin{equation} B = \frac{e}{\kappa}\int^x \bar \psi \psi \,dx'.\label{B} \end{equation}

From another side, flatness of connection (\ref{CSRE2}) implies $A_0 = \partial_0 \phi$,
$A_1 = \partial_1 \phi$, and these potentials can be removed by the gauge transformation,
$\psi = e^{-e \phi} \Psi $.
As a result we obtain the Schr\"odinger  equation with
self-interacting nonlinear potential $V(\bar\psi \psi)$,
and the quantum potential

\begin{equation}i\partial_0 \Psi + \partial_1^{2}\Psi + V'\Psi = s \frac{\partial^2_1|\Psi|}{|\Psi|}\Psi \label{NSEV}\end{equation}

\subsection{Madelung representation and RNLS}

If we substitute the Madelung Ansatz
$\Psi = \sqrt{\rho} \,e^{-i S}$
to wave function in (\ref{NSEV})
then we get the coupled system

\begin{equation} \partial_0 S - (\partial_1 S)^2 + V'(\rho) + (1-s) \frac{\partial^2_1 \sqrt{\rho}}{\sqrt{\rho}} = 0 \label{HJ}\end{equation}

\begin{equation} \partial_0 \rho - \partial_1 (2 \rho \,\partial_1 S) = 0 \label{CE} \end{equation}
For velocity field $v = -2 \partial_1 S$ it implies the hydrodynamical system

\begin{equation} \partial_0 v + v \partial_1 v = 2 \partial_1\left( V'(\rho) + (1-s) \frac{\partial^2_1\sqrt{\rho}}{\sqrt{\rho}}\right) \label{M1} \end{equation}

\begin{equation} \partial_0 \rho + \partial_1 (\rho\, v) = 0 \label{M2} \end{equation}
which is the Madelung fluid representation of (\ref{NSEV}).

First we consider the under-critical case, when the strength of the quantum potential $s < 1$.
Then introducing rescaled the time and the phase
$ \tilde t = t \sqrt{1-s}$, $\tilde S = \frac{S}{\sqrt{1-s}}$,
for new wave function $\tilde\Psi = \sqrt{\rho} \,e^{-i \tilde S}$ we get

\begin{equation}i\partial_{\tilde 0} \tilde\Psi + \partial_1^{2}\tilde\Psi + \frac{V'}{1-s}\tilde\Psi = 0\label{NSEVT}\end{equation}

In the over-critical case when $s>1$, we can't reduce (\ref{NSEV}) to (\ref{NSEVT}).
However, if   we introduce pair of real functions

\begin{equation} e^{(+)}(x,t) = \sqrt{\rho} e^{\tilde S},\,\,\,\,\,
e^{(-)}(x,t) = \sqrt{\rho} e^{-\tilde S}\label{EPM} \end{equation}
instead of one complex wave function,
then we get the time reversal pair of reaction-diffusion equations

\begin{equation}\partial_{\tilde 0} e^{(+)} + \partial_1^{2} e^{(+)} - \frac{V'}{s-1} e^{(+)} = 0\label{RDP}\end{equation}

\begin{equation}-\partial_{\tilde 0} e^{(-)} + \partial_1^{2} e^{(-)} - \frac{V'}{s-1} e^{(-)} = 0\label{RDM}\end{equation}
where $\tilde t = t \sqrt{s-1}$, $\tilde S = \frac{S}{\sqrt{s-1}}$, $\rho = e^{(+)}e^{(-)}$, $V  = V (e^{(+)}e^{(-)})$.

If interaction between material particles is the delta function pair form then potential
$V(\rho) = g \rho^2/2$ and equation (\ref{NSEV}) becomes

\begin{equation}i\partial_0 \Psi + \partial_1^{2}\Psi + g |\Psi|^2\Psi = s \frac{\partial^2_1|\Psi|}{|\Psi|}\Psi \label{RNLS}\end{equation}

We called this equation  the Resonant Nonlinear Schrodinger equation (RNLS).
It appears in the study of low-dimensional gravity model on a line, the Jackiw-Teitelboim model \cite{Pashaev et al.a}, and in description of cold plasma \cite{Lee et al.a}. For under-critical case $s<1$ it reduces to the standard NLS equation (\ref{NSEVT}), and is integrable model. For the over-critical case $s>1$ it reduces to the couple of cubic  reaction-diffusion equations (\ref{RDP}), (\ref{RDM}), which is also integrable as SL(2,R) NLS from the AKNS hierarchy. In the last case new resonance phenomena for envelope solitons take place \cite{Pashaev et al.a}. In the next section we will discuss another reduction of Chern-Simons theory with dynamical field $B$, and show that in this case resonant versions of the JNLS and DNLS equations appear.

\section{Dynamical BF theory}

To do the gauge field component $B$ in Section 2 to be dynamical, following \cite{Aglietti} we introduce the corresponding kinetic term so that

\begin{equation} \begin{array}{cc} L = \kappa B ( \partial_0 A_{1} - \partial_{1} A_{0}) + \theta \,\partial_0 B \partial_1 B + {i \over 2}(\bar\psi (\partial_0 + i e A_0)\psi \\  \\  - \psi (\partial_0 - i e  A_0)\bar \psi) - (\partial_1 - ie A_1)\bar\psi (\partial_1 + ie A_1)\psi + s\partial_1 |\psi| \partial_1 |\psi| + V(\bar\psi \psi),\end{array} \label{DBF}\end{equation}
Then equations of motion are

\begin{equation}i(\partial_0 + ie  A_0)\psi + (\partial_1 + ie A_1)^{2}\psi + V'\psi = s \frac{\partial^2_1|\psi|}{|\psi|}\psi, \label{DCSRE1}\end{equation}

\begin{equation} \partial_0 A_1 - \partial_1 A_0 = \frac{2\theta}{\kappa} \partial_0 \partial_1 B \label{DCSRE2}\end{equation}

\begin{equation} \partial_1 B = \frac{e}{\kappa} \rho \label{DCSRE3}\end{equation}

\begin{equation} \partial_0 B = - \frac{e}{\kappa} J \label{DCSRE4}\end{equation}
where the particle and the momentum density are

\begin{equation} \rho =  \bar \psi \psi,\,\,\, J = \frac{1}{i} [\bar\psi (\partial_1 + ieA_1)\psi - \psi (\partial_1 - ie A_1)\bar\psi]  = j + 2e A_1 \rho\end{equation}
and
$j = -i[\bar\psi \partial_1 \psi - \psi \partial_1 \bar\psi]$.
Equation (\ref{DCSRE1}) implies the conservation law

\begin{equation} \partial_0 \rho + \partial_1 J = 0\label{CL} \end{equation}
This conservation law is the compatibility condition for the
system (\ref{DCSRE3}),(\ref{DCSRE4}). and allows us to write

\begin{equation} \partial_0 \partial_1 B = \frac{e}{\kappa}[\,\alpha \partial_0\rho - (1-\alpha) \partial_1 J\,] \label{BAE}\end{equation}
where $\alpha$ is an arbitrary real constant.
Substituting (\ref{BAE}) to (\ref{DCSRE2}) and combining terms
under the same derivatives we have

\begin{equation} \partial_0 \left(A_1 - \frac{2\theta e}{\kappa^2}\alpha \rho\right) - \partial_1 \left( A_0 - \frac{2\theta e}{\kappa^2}(1-\alpha) J\right) = 0 \label{Aequation}\end{equation}
The system (\ref{DCSRE1})-(\ref{DCSRE4}) is invariant under the local $U(1)$ gauge transformations

\begin{equation}
\psi \rightarrow \psi' = e^{-ie \phi(x,t)} \psi,\,\,\,\,
A_\mu \rightarrow A'_\mu = A_\mu + \partial_\mu \phi
\end{equation}
Then solving (\ref{Aequation}) we have

\begin{equation} A_1 = \frac{2\theta e}{\kappa^2}\alpha \rho + \partial_1 \phi,\,\,\, A_0 = \frac{2\theta e}{\kappa^2}(1-\alpha) J + \partial_0\phi \label{Apot}\end{equation}
and for the gauge invariant field $\Psi = e^{i e \phi} \psi$ it gives

\begin{equation}i\left(\partial_0 + i \frac{2\theta e^2}{\kappa^2}(1-\alpha)J\right)\Psi + \left(\partial_1 + i \frac{2\theta e^2}{\kappa^2}\alpha \rho\right)^{2}\Psi + V'\Psi = s \frac{\partial^2_1|\psi|}{|\Psi|}\Psi, \label{DynamNLS}\end{equation}
where $J = j + \frac{4\theta e^2}{\kappa^2}\alpha \rho^2$, $j = -i[\bar\Psi \Psi_x - \Psi \bar\Psi_x]$, $\rho = \bar\Psi \Psi$.
Finally we have

\begin{equation} \begin{array}{cc} i \Psi_t + \Psi_{xx} + i\frac{2\theta e^2}{\kappa^2}\left[(2\alpha +1)|\Psi|^2 \Psi_x + (2\alpha -1)\Psi^2 \bar\Psi_x
 \right]
\\ \\ + 4 \frac{\theta^2 e^4}{\kappa^4}\alpha (\alpha -2) |\Psi|^4 \Psi + V' \Psi = s \frac{|\Psi|_{xx}}{|\Psi|}\Psi \end{array}\label{adnls}\end{equation}
where partial differentiation notations are evident.
The remaining gauge transformation for this equation is just the global
$U(1)$ transformation: $\Psi \rightarrow e^{i\lambda} \Psi$, $\lambda = const.$

\subsection{Reductions of general RDNLS}

The behavior of equation   (\ref{adnls})  depends on value of parameter $s$.
If we replace $\Psi = e^{R - iS}$
then we have couple of equations

\begin{eqnarray}
R_t - (S_{xx} + 2 R_x S_x) + \frac{2\theta e^2}{\kappa^2} 4\alpha  R_x e^{2R} = 0,\nonumber
\\
S_t - S^2_x + (1-s) (R_{xx} + R_x^2) + \frac{2\theta e^2}{\kappa^2} 2 S_x e^{2R} + \frac{4\theta^2 e^4}{\kappa^4}
\alpha (\alpha - 2) e^{4R} + V' =0,\nonumber
\end{eqnarray}
determining the Madelung fluid representation

\begin{eqnarray}
\rho_t + (\rho v + \frac{2\theta e^2}{\kappa^2} 2\alpha \rho^2)_x = 0,
\\
v_t + v v_x = 2 \left[ (1-s) \frac{\sqrt{\rho}_{xx}}{\sqrt{\rho}} - \frac{2\theta e^2}{\kappa^2}\rho v
+ \left(\frac{2\theta e^2}{\kappa^2}\right)^2 \alpha (\alpha-2) \rho^2 + V'\right]_x
\end{eqnarray}

For $s < 1$ for redefined variables $t \sqrt{1-s} \equiv \tilde t$,
 $S/\sqrt{1-s} \equiv \tilde S$, $\tilde \Psi \equiv e^{R - i \tilde S}$,
 we have

\begin{equation} \begin{array}{cc} i \tilde\Psi_{\tilde t} + \tilde\Psi_{xx} + i\frac{2\theta e^2}{\kappa^2 \sqrt{1-s}}\left[(2\alpha +1)|\tilde\Psi|^2 \tilde\Psi_x + (2\alpha -1)\tilde\Psi^2 \bar{\tilde\Psi}_x
 \right]
\\ \\ + 4 \frac{\theta^2 e^4}{\kappa^4 (1-s)}\alpha (\alpha -2) |\tilde\Psi|^4 \tilde\Psi + \frac{V'}{1-s} \tilde\Psi = 0 \end{array}\label{gdnls}\end{equation}

Similar to the Chern-Simons 2+1 dimensional case \cite{Pashaev et al.c} we have effective result of the quantum potential in the rescaling of the statistical parametr $\kappa^2 \rightarrow \kappa^2 \sqrt{1-s}$, but in contrast no quantization of this parameter now appears. Transformation between wave functions has nonlinear form

\begin{equation}
\Psi(x,t) = |\tilde\Psi|\left(\frac{\tilde\Psi}{|\tilde\Psi|}\right)^{\sqrt{1-s}} (x, t \sqrt{1-s})
\end{equation}

For $s > 1$ it is impossible to reduce the system to the Schrodinger type form. However for redefined parameters
$t \sqrt{s-1} \equiv \tilde t$, $S/\sqrt{s-1} \equiv \tilde S$
and two real functions
$E^+ = e^{R + \tilde S}$, $E^- = e^{R - \tilde S}$
we get

\begin{equation} \begin{array}{cc} \mp E^{\pm}_{\tilde t} + E^{\pm}_{xx} \mp\frac{2\theta e^2}{\kappa^2 \sqrt{s-1}}\left[(2\alpha +1) E^+ E^- E^{\pm}_x + (2\alpha -1){E^{\pm}}^2 E^{\mp}_x
 \right]
\\ \\ - 4 \frac{\theta^2 e^4}{\kappa^4 (s-1)}\alpha (\alpha -2) (E^{+}E^-)^2 E^{\pm} - \frac{V'}{s-1} \tilde E^{\pm} = 0 \end{array}\label{rddnls}\end{equation}

\subsection{Gauge transformation}

 We notice that in the gauge potential representation (\ref{Apot}) ,
 the gauge function
 $\phi = \phi^{(\alpha)}$ depends on $\alpha$:

\begin{equation} A_1 = \frac{2\theta e}{\kappa^2}\alpha \rho + \partial_1 \phi^{(\alpha)},\,\,\, A_0 = \frac{2\theta e}{\kappa^2}(1-\alpha) J + \partial_0\phi^{(\alpha)} \label{ApotA}\end{equation}

Comparison with the case $\alpha = 0$

\begin{equation} A_1 = \partial_1 \phi^{(0)},\,\,\, A_0 = \frac{2\theta e}{\kappa^2} J + \partial_0\phi^{(0)} \label{Apot0}\end{equation}
gives relations

\begin{equation} \partial_1 (\phi^{(0)} - \phi^{(\alpha)}) =  \frac{2\theta e}{\kappa^2}\alpha \,\rho,\,\,\, \partial_0 (\phi^{(0)} - \phi^{(\alpha)}) =  -\frac{2\theta e}{\kappa^2}\alpha \,J \label{0A}\end{equation}

Compatibility of this system  is ensured by the continuity equation (\ref{CL}).
Then corresponding gauge transformed wave functions $\Psi^{(\alpha)}$ and $\Psi^{(0)}$

\begin{equation} \psi = e^{-i e \phi^{(\alpha)}} \Psi^{(\alpha)} = e^{-i e \phi^{(0)}} \Psi^{(0)} \end{equation}
are related by

\begin{equation} \Psi^{(\alpha)} = e^{-i e (\phi^{(0)}- \phi^{(\alpha)})} \Psi^{(0)} \label{GTA0} \end{equation}

Integrating (\ref{0A}) and substituting to (\ref{GTA0}) we have gauge transformation between equations (\ref{adnls}) and the same equation with $\alpha = 0$:

\begin{equation} \Psi^{(\alpha)} = \exp \left(-i \frac{2\theta e^2}{\kappa^2}\alpha \int^x \rho dx'\right)\, \Psi^{(0)} \label{GT1A0} \end{equation}

From this relation we can connect two samples of equation   (\ref{adnls}) with different constants
$\alpha$ and $\beta$

\begin{equation} \Psi^{(\alpha)} = \exp \left(-i \frac{2\theta e^2}{\kappa^2}(\alpha - \beta) \int^x \rho dx'\right)\, \Psi^{(\beta)} \label{GT1A12} \end{equation}
Indeed one can check easily from

\begin{equation}
\bar\Psi^{(\alpha)} \Psi^{(\alpha)} = \bar\Psi^{(\beta)} \Psi^{(\beta)}, \,\,\,
\bar\Psi^{(\alpha)} (\partial_1 + i\nu \alpha \rho)\Psi^{(\alpha)} = \bar\Psi^{(\beta)} (\partial_1 + i\nu \alpha \rho)\Psi^{(\beta)}
\end{equation}
that $\rho^{(\alpha)} = \rho^{(\beta)}$, $J^{(\alpha)} = J^{(\beta)}$, and

\begin{equation}
(\partial_1 + i\nu \alpha \rho)\Psi^{(\alpha)} = e^{-i\nu(\alpha - \beta) \int^x \rho dx'}
(\partial_1 + i\nu \beta \rho)\Psi^{(\beta)}
\end{equation}

\begin{equation}
(\partial_0 + i\nu (1-\alpha) J)\Psi^{(\alpha)} = e^{-i\nu(\alpha - \beta) \int^x \rho dx'}
(\partial_0 + i\nu (1-\beta) J)\Psi^{(\beta)}
\end{equation}
where $\nu \equiv \frac{2\theta e^2}{\kappa^2}$

The gauge transformation (\ref{GT1A12}) for the Madelung representation implies

\begin{equation}
S^{(\alpha)} - S^{(\beta)} = \nu (\alpha-\beta) \int^x \rho dx' + 2\pi n, \,\,\,
R^{(\alpha)} = R^{(\beta)}
\end{equation}
For $s<1$ it gives $U(1)$ gauge transformation for (\ref{gdnls}) in the form

\begin{equation}
\tilde\Psi^{(\alpha)} = \tilde\Psi^{(\beta)} e^{-i\frac{\nu}{\sqrt{1-s}}(\alpha-\beta)\int^x \rho dx'} e^{-i\frac{2\pi n}{\sqrt{1-s}}} \label{GT1}
\end{equation}
The last multiplier can be absorbed by the global phase transformation on $\Psi$.

For $s >1$ the above $U(1)$ gauge transformation give rise to the local $SO(1,1)$ scale transformation (the Weyl transformation) for equation (\ref{rddnls})

\begin{equation}
E^{\pm(\alpha)} = E^{\pm(\beta)} e^{\pm\frac{\nu}{\sqrt{s-1}}(\alpha-\beta)\int^x \rho dx'} e^{\pm\frac{2\pi n}{\sqrt{s-1}}}
\end{equation}

\section{Integrable DRD Systems}

It was shown above that the one dimensional problem of anyons in quantum potential
with a specific form of the three-body interaction, can be reduced to the general resonant DNLS equation.

\subsection{General Resonant DNLS}

This equation
\begin{eqnarray} i \Psi_{\tilde t} + \Psi_{xx} + i\tilde \nu\left[(2\alpha +1)|\Psi|^2 \Psi_x + (2\alpha -1)\Psi^2 \bar\Psi_x
 \right] \nonumber
 \\+ 4 \tilde\nu^2(\alpha - \frac{1}{2}) (\alpha -\frac{3}{2}) |\Psi|^4 \Psi  = s \frac{|\Psi|_{xx}}{|\Psi|}\Psi \label{generalrdnls}\end{eqnarray}
is integrable for any values of parameter $\alpha$.

\subsection{The Resonant Case}

For special case $s > 1$, by the Madelung transformation $\Psi = e^{R - iS}$ and introduction of two new
real functions $E^{+} = e^{R + S}$, $E^{-} = e^{R-S}$ we get the general DRD system

\begin{equation} \begin{array}{cc} \mp E^{\pm}_{t} + E^{\pm}_{xx} \\\mp\frac{2\theta e^2}{\kappa^2 \sqrt{s-1}}\left[(2\alpha +1) E^+ E^- E^{\pm}_x + (2\alpha -1){E^{\pm}}^2 E^{\mp}_x
 \right]
\\ \\ - 4 \frac{\theta^2 e^4}{\kappa^4 (s-1)}(\alpha - \frac{1}{2}) (\alpha - \frac{3}{2}) (E^{+}E^-)^2 E^{\pm}  = 0, \end{array}\label{gdrd}\end{equation}
where $\theta$ is the statistical parameter.  

This system has particular reductions

1. DRD-I, ($\alpha = 3/2$)
\begin{eqnarray} - E^{+}_{ t} + E^{+}_{xx} - 2\nu (E^+ E^- E^{+})_x &=& 0, \\
+ E^{-}_{ t} + E^{-}_{xx} + 2 \nu (E^+ E^- E^{-})_x &=& 0\label{drd1}\end{eqnarray}
2. DRD-II, ($\alpha = 1/2$)
\begin{eqnarray} - E^{+}_{ t} + E^{+}_{xx} - 2\nu E^+ E^- E^{+}_x &=& 0, \\
+ E^{-}_{ t} + E^{-}_{xx} + 2 \nu E^+ E^- E^{-}_x &=& 0\label{drd2}\end{eqnarray}
3. DRD-III, ($\alpha = -1/2$)
\begin{eqnarray} - E^{+}_{ t} + E^{+}_{xx} + 2 \nu {E^{+}}^2 E^{-}_x - 2  \nu^2 (E^{+}E^-)^2 E^{+}  &=& 0 ,\\
+ E^{-}_{ t} + E^{-}_{xx} - 2 \nu {E^{-}}^2 E^{+}_x - 2  \nu^2 (E^{+}E^-)^2 E^{-}  &=& 0
\label{drd3}\end{eqnarray}
4. JRD, ($\alpha = 0$)
\begin{eqnarray}\mp E^{\pm}_{t} + E^{\pm}_{xx} -\nu\left[ E^+_x E^-   - E^{+} E^{-}_x
 \right] E^\pm \nonumber \\- \frac{3\nu^2}{4} (E^{+}E^-)^2 E^{\pm}  = 0 \label{jdrd}\end{eqnarray}

\section{Resonant Hydrodynamic Systems}

To find hydrodynamic form of the above equations we introduce velocity variables
according to the Cole-Hopf transformation
\begin{equation}
v^+ = (\ln E^+)_x,\,\,\, v^- = (\ln E^-)_x
\end{equation}
and density
\begin{equation}
\rho = E^+ E^-.
\end{equation}
Then by identity
\begin{equation}
\rho_x = \rho v^+ + \rho v^-,
\end{equation}
we can rewrite the DRD system in a closed form for only one of the couples of hydrodynamic variables $(\rho, v^+)$ or $(\rho, v^-)$.

\subsection{Hydrodynamic Form for DRD-I}

For DRD-I case it gives the new hydrodynamic system
\begin{eqnarray}
v^+_t = [v^+_x + (v^+)^2 - 2 \nu (\rho_x + \rho v^+)]_x , \nonumber\\
\rho_t + \rho_{xx} = [2 \rho v^+ - 3 \nu \rho^2 ]_x .\label{BKDRD1}
\end{eqnarray}

\subsection{Hydrodynamic Form for DRD-II}

For DRD-II case first we get the coupled heat equation with transport
\begin{eqnarray}
-E^+_t + E^+_{xx} - 2\nu \rho E^+_x = 0 , \nonumber\\
\rho_t + \rho_{xx}= (2 \rho (\ln E^+)_x - \nu \rho^2)_x .
\end{eqnarray}
Then the hydrodynamic form for this system is
\begin{eqnarray}
v^+_t = [v^+_x + (v^+)^2 - 2 \nu \rho v^+]_x ,   \nonumber\\
\rho_t + \rho_{xx} = [2 \rho v^+ -  \nu \rho^2 ]_x .\label{BKDRD2}
\end{eqnarray}

\subsection{Hydrodynamic Form for DRD-III}

For DRD-III case it gives the new hydrodynamic system
\begin{eqnarray}
v^+_t = [v^+_x + (v^+)^2 + 2 \nu (\rho - \rho v^+)- 2 \nu^2 \rho^2 ]_x  ,\nonumber\\
\rho_t + \rho_{xx} = [2 \rho v^+ +  \nu \rho^2 ]_x .\label{BKDRD3}
\end{eqnarray}

\subsection{Hydrodynamic Form for JRD}

For JRD case it gives the new hydrodynamic system
\begin{eqnarray}
v^+_t = [v^+_x + (v^+)^2 - \nu (2\rho v^+ - \rho_x)- \frac{3}{4} \nu^2 \rho^2 ]_x  ,\nonumber\\
\rho_t + \rho_{xx} = [2 \rho v^+]_x .
\end{eqnarray}
In all above cases for $v^-$ we have the  system with replaced $t \rightarrow -t$, $\nu \rightarrow -\nu$.

\subsection{Generic Case}

For the generic case of arbitrary $\alpha$ firstly we have the system
\begin{eqnarray}
-E^+_t + E^+_{xx}- \nu [2\rho E^+_x + (2\alpha -1)\rho_x E^+] \nonumber \\-\nu^2 (\alpha - \frac{1}{2})(\alpha - \frac{3}{2}) \rho^2 E^+ = 0 ,
\nonumber \\
\rho_t = [2 \rho (\ln E^+)_x - \rho_x - 2 \nu \alpha \rho^2]_x .\label{genericDRD}
\end{eqnarray}

 It gives the new hydrodynamic system
\begin{eqnarray}
v^+_t = [v^+_x + (v^+)^2 - \nu (2\rho v^+ + (2\alpha - 1)\rho_x) \nonumber
\\-  \nu^2 (\alpha-\frac{1}{2})(\alpha - \frac{3}{2}) \rho^2 ]_x  ,\nonumber\\
\rho_t + \rho_{xx} = [2 \rho v^+ - 2\nu \alpha \rho^2]_x .\label{genericcase}
\end{eqnarray}

\section{RNLS and Broer-Kaup system}

The RNLS for $s > 1$ can be transformed to the reaction-diffusion system
\begin{eqnarray}
 R^{+}_t &=& R^{+}_{xx} + 2\nu R^+ R^- R^{+},\\
 -R^{-}_t &=& R^{-}_{xx} + 2\nu R^+ R^- R^{-} .\label{RD}
\end{eqnarray}
By substitution $v^+ = (\ln E^+)_x, \,\,\,\rho = E^+ E^-$,
it can be transformed to the
the hydrodynamic form as the Broer-Kaup system, \cite{Broer}, \cite{Kaup},
\begin{eqnarray}
v^+_t = (v^+_x + (v^+)^2)_x + 2 \nu \rho_x,  \nonumber\\
\rho_t + \rho_{xx} = (2 \rho v^+)_x .\label{BKP}
\end{eqnarray}
If $v^- = (\ln E^-)_x, \,\,\,\rho = E^+ E^-$, then we have
\begin{eqnarray}
-v^-_t = (v^-_x + (v^-)^2)_x + 2 \nu \rho_x,  \nonumber\\
-\rho_t + \rho_{xx} = (2 \rho v^-)_x .\label{BKM}
\end{eqnarray}

\section{Relation with Broer-Kaup System}

Given $E^+(x,t)$, $E^-(x,t)$ satisfying general DRD system (\ref{gdrd}), then real functions

\begin{eqnarray}
R^+ &=& E^+ e^{-(\alpha+ \frac{1}{2})\nu \int^x E^+ E^- } ,\nonumber \\
 R^- &=& \left[E^-_x + (\alpha - \frac{1}{2})\nu E^+ E^- E^+ \right] e^{(\alpha+ \frac{1}{2})\nu \int^x E^+ E^- }
 \end{eqnarray}
or

\begin{eqnarray}
 R^+ &=& \left[-E^+_x + (\alpha - \frac{1}{2})\nu E^+ E^- E^+ \right] e^{-(\alpha+ \frac{1}{2})\nu \int^x E^+ E^- },\nonumber \\
R^- &=& E^- e^{(\alpha+ \frac{1}{2}) \nu \int^x E^+ E^- }
\end{eqnarray}
satisfy the reaction-diffusion (RD) system
\begin{eqnarray}
 R^{+}_t &=& R^{+}_{xx} + 2\nu R^+ R^- R^{+},\\
 -R^{-}_t &=& R^{-}_{xx} + 2\nu R^+ R^- R^{-} .\label{RD1}
\end{eqnarray}
From this fact we can get next result.

If $v^+_E$ and $\rho_E$ satisfy (\ref{genericcase}) then $v^+_R$ and $\rho_R$ determined by
\begin{eqnarray}
v^+_R &=& v^+_E - (\alpha + \frac{1}{2})\nu \rho_E, \\
\rho_R &=& (\rho_E)_x - \rho_E v^+_E + (\alpha - \frac{1}{2}) \nu \rho^2_E,
\end{eqnarray}
is solution of the Broer-Kaup system (\ref{BKP}).
 For $v^-_E$ and $\rho_E$  satisfying the analog of system (\ref{genericcase}),
\begin{eqnarray}
v^-_R &=& v^-_E + (\alpha + \frac{1}{2})\nu \rho_E + [\ln (v^-_E + (\alpha - \frac{1}{2})) \nu \rho_E]_x, \\
\rho_R &=& \rho_E v^-_E + (\alpha - \frac{1}{2}) \nu \rho^2_E
\end{eqnarray}
is solution of (\ref{BKM}).

Similar way we can get result.

If $v^+_E$ and $\rho_E$ satisfy (\ref{genericcase}) then $v^+_R$ and $\rho_R$ determined by

\begin{eqnarray}
v^+_R &=& v^+_E - (\alpha + \frac{1}{2})\nu \rho_E + [\ln (-v^+_E + (\alpha - \frac{1}{2})) \nu \rho_E]_x, \\
\rho_R &=& -\rho_E v^+_E + (\alpha - \frac{1}{2}) \nu \rho^2_E
\end{eqnarray}
is solution of the Broer-Kaup system (\ref{BKP}).
 For $v^-_E$ and $\rho_E$  satisfying the analog of system (\ref{genericcase}),
\begin{eqnarray}
v^-_R &=& v^-_E + (\alpha + \frac{1}{2})\nu \rho_E, \\
\rho_R &=& -(\rho_E)_x + \rho_E v^-_E + (\alpha - \frac{1}{2}) \nu \rho^2_E
\end{eqnarray}
is solution of (\ref{BKM}).

\section{B\"acklund Transformation}

When $\rho \equiv 0$, both systems (\ref{genericcase}) and (\ref{BKP}) reduce to the Burgers
equation. Then the above Miura type transformations reduce to the auto-B\"acklund transformations
\begin{equation}
v^+_R = v^+_E + (\ln v^+_E)_x,\,\,\,v^-_R = v^-_E + (\ln v^-_E)_x
\end{equation}
for the Burgers and anti-Burgers equations correspondingly.

\section{Classical Bousinesque Systems}

If in (\ref{BKP}) we change variables
\begin{equation}
p^+ = v^+_x + 2 \nu \rho,
\end{equation}
then we get the classical Bousinesque system
\begin{eqnarray}
v^+_t &=& ((v^+)^2 + p^+)_x ,\\
p^+_t &=& (v^+_{xx} + 2 p^+ v^+)_x .
\end{eqnarray}
Similar way in (\ref{BKM}) by variable change
\begin{equation}
p^- = v^-_x + 2 \nu \rho
\end{equation}
we get
\begin{eqnarray}
-v^-_t &=& ((v^-)^2 + p^-)_x ,\\
-p^-_t &=& (v^-_{xx} + 2 p^- v^-)_x .
\end{eqnarray}

\section{Bilinear Form and Solitons}

By substitution $E^{\pm} = g^{\pm}/f^{\pm}$ to (\ref{genericDRD})
we have bilinear
representation
 \begin{eqnarray} (\mp D_{\tilde t} + D^2_{x})(g^{\pm} \cdot f^{\pm}) = 0, \label{bl1} \\
 D^2_{x} (f^{+} \cdot f^{-}) + {1 \over 2} D_{x} (g^{+} \cdot g^{-}) = 0, \\
 D_{x} (f^{+} \cdot f^{-}) + \alpha  g^{+} g^{-} = 0, \label{bl3} \end{eqnarray}
where $\alpha = \frac{1}{2}$ ( DRD-II case), or $\alpha = -\frac{1}{2}$  (DRD-III case).
We note that only in these two cases the Hirota substitution has simple bilinear form.
Then for solution of the hydrodynamics systems  (\ref{BKDRD2}) and (\ref{BKDRD3}) we have
\begin{equation}
v^+ = (\ln E^+)_x = \frac{g^+_x}{g^+}-\frac{f^+_x}{f^+},\end{equation}
\begin{equation}\rho = E^+ E^- = (\ln \frac{f^+}{f^-})_x .
\end{equation}
Bilinearization for arbitrary $\alpha$ can be derived by the gauge transformation, so that
\begin{equation}
{E^+} =  \frac{g^+}{(f^+)^{\frac{1}{2} +\alpha} (f^-)^{\frac{1}{2} - \alpha}},
{E^-} =  \frac{g^-}{(f^+)^{\frac{1}{2} -\alpha} (f^-)^{\frac{1}{2} + \alpha}}
\end{equation}
It implies next substitution for equation (\ref{genericcase})
\begin{equation}
v^+ = (\ln E^+)_x = \frac{g^+_x}{g^+}-(\frac{1}{2} + \alpha)\frac{f^+_x}{f^+} - (\frac{1}{2} - \alpha)\frac{f^-_x}{f^-},
\end{equation}
\begin{equation}
\rho = E^+ E^- = (\ln \frac{f^+}{f^-})_x .
\end{equation}

\subsection{One Soliton Solution}
For one soliton solution we have

\begin{equation}g^{\pm} = e^{\eta_1^{\pm}},\,\, f^{\pm} = 1 + e^{\phi_{11}^{\pm}} e^{\eta_1^{+} + \eta_1^{-}},\end{equation}
where $e^{\phi_{11}^{\pm}} = \mp \frac{k_1^{\pm}}{(k^{+}_1 + k^{-}_1)^2}$,
$\eta^{\pm}_1 = k^{\pm}_1 x \pm (k^{\pm}_1)^2  t + \eta^{\pm(0)}_{1}$.
For regularity of this solution we choose conditions $k_1^{-} > 0$ and $k_1^{+} < 0$, then $-\tilde v < k < \tilde v$, where
$k = k_1^+ + k_1^-$, $\tilde v = k_1^- - k_1^+ $, $- k x_0^{\pm} = \eta_1^{+(0)}+ \eta_1^{-(0)}
+ \phi_{11}^{\pm}$. Then
velocity is positive
$\tilde v >0$, so that our dissipaton is chiral. For the density we have soliton solution
\begin{equation}
\rho = E^+ E^- = \frac{ k^2}{\sqrt{\tilde v^2 - k^2} \cosh k(x - \tilde v  t - x_0) +\tilde v }
\end{equation}
where $2 x_0 = x_0^+ + x_0^- $,
and for velocity field
\begin{equation}
v^+ = \frac{k^+_1 - k^-_1 e^{\phi^+_{11}} e^{\eta^+_1 + \eta^-_1}}{1 + e^{\phi^+_{11}} e^{\eta^+_1 + \eta^-_1}},
\end{equation}
the kink solution
\begin{equation}
v^+ = -\frac{\tilde v }{2} - \frac{k}{2} \tanh \frac{k}{2}(x - \tilde v  t - x_0) .
\end{equation}

\subsection{Integrals of Motion}
The particle number, momentum and energy  integrals are given respectively
\begin{equation}
N = \int^\infty_{-\infty} \rho dx = - \frac{1}{\nu}\ln \frac{f^+}{f^-}|^\infty_{-\infty}\label{nintegral}
\end{equation}
\begin{equation}
P = - \int^{\infty}_{-\infty} \rho v^+ dx =\frac{1}{2\nu} \ln(f^+ f^-)_x|^\infty_{-\infty}
\end{equation}
\begin{equation}
E = -\int^\infty_{-\infty}[ \rho (v^+)^2 - \rho_x v^+ - \nu \rho^2 v^+ ]dx .
\end{equation}
Then substituting for one soliton solution we find
\begin{equation}
N = \frac{1}{ \nu}\ln \frac{\tilde v + |k|}{\tilde v - |k|},\,\,\,P = \frac{|k|}{ \nu},\,\,\,
E =  \frac{\tilde v |k|}{2\nu}.\label{1disintegral}
\end{equation}
The mass of soliton $M = |k|/( \nu \tilde v)$ in terms of particle number becomes
$
M = \frac{1}{ \nu} \tanh \frac{N  \nu}{2}$, and
for the momentum and the energy we have non-relativistic free particle form
$P = M \tilde v,\,\,\,\,E = \frac{M \tilde v^2}{2}$.

For the process of fusion or fission of two solitons then the next conditions should be valid
\begin{equation}
N = N_1 + N_2,\,\,\,P = P_1 + P_2,\,\,\,E = E_1 + E_2
\end{equation}
Using   (\ref{1disintegral}) after some algebraic manipulations we get the resonance condition
\begin{equation}
|\tilde v_1 - \tilde v_2| = |k_1| + |k_2|\label{rescond}
\end{equation}
where $\tilde v_a = k_a^- - k_a^+$, $k_a = k_a^- + k_a^+$, $a=1,2$.

\subsection{Two Soliton Solution}

For two soliton solution we have
\begin{equation}
g^{\pm} = e^{\eta^{\pm}_{1}} + e^{\eta^{\pm}_{2}} + \alpha^{\pm}_{1}e^{\eta^{+}_{2} + \eta^{-}_{2} + \eta^{\pm}_{1}} + \alpha^{\pm}_{2}e^{\eta^{+}_{1} + \eta^{-}_{1} + \eta^{\pm}_{2}},
\label{eq:2s1}\end{equation}
\begin{equation}
f^{\pm} = 1 + \sum_{i,j = 1}^2 e^{\phi^{\pm}_{ij}} e^{\eta^{+}_{i} + \eta^{-}_{j}}
+ \beta^{\pm}e^{\eta^{+}_{1} + \eta^{-}_{1} + \eta^{+}_{2} + \eta^{-}_{2}},
\label{eq:2s2}\end{equation}
where $\eta^{\pm}_{i} = k^{\pm}_{i}x \pm (k^{\pm}_{i})^2  t + \eta^{\pm}_{i0}$, $k^{nm}_{ij} \equiv (k^{n}_{i} + k^{m}_{j})$ and
\begin{equation}
\alpha^{\pm}_{1} = \pm {1 \over 2}{k^{\mp}_{2} (k^{\pm}_{1} - k^{\pm}_{2})^2\over (k^{+-}_{22})^2(k^{\pm\mp}_{12})^2 },\,\,\,
\alpha^{\pm}_{2} = \pm {1 \over 2}{k^{\mp}_{1} (k^{\pm}_{1} - k^{\pm}_{2})^2\over (k^{+-}_{11})^2(k^{\pm\mp}_{21})^2 },
\label{eq:2s3}\end{equation}
\begin{equation}
\beta^{\pm} = {(k^{+}_{1} - k^{+}_{2})^2 (k^{-}_{1} - k^{-}_{2})^2 \over 4(k^{+-}_{11}k^{+-}_{12}k^{+-}_{21}k^{+-}_{22})^2}k^{\pm}_{1}k^{\pm}_{2},
\label{eq:2s4}\end{equation}
\begin{equation}
e^{\phi^{\pm}_{ii}} = \mp {k^{\pm}_{i} \over 2 (k^{+-}_{ii})^2},\,\,\,
e^{\phi^{+}_{ij}} = {-k^{+}_{i}  \over 2 (k^{+-}_{ij})^2},
\,\,\,
e^{\phi^{-}_{ij}} =  { k^{-}_{j} \over 2 (k^{+-}_{ij})^2}.
\label{eq:2s5}\end{equation}
By regularity we have
$k^{+}_{i} \leq 0$, $k^{-}_{i} \geq 0$ in the Case 1, and
$k^{+}_{i} \geq 0$, $k^{-}_{i} \leq 0$ in the Case 2.
Then solving the resonance condition (\ref{rescond}) we find that for every
solution of this algebraic equation, the coefficient $\beta$ vanishes or becomes infinite.
In both cases two soliton solution reduces to the one soliton solution.
Hence the solution describes a collision of two solitons propagating in the same direction and  at some
value of parameters creating the resonance states (see Fig.1 and Fig.2).

\begin{figure}[h]
\begin{center}
\epsfig{figure=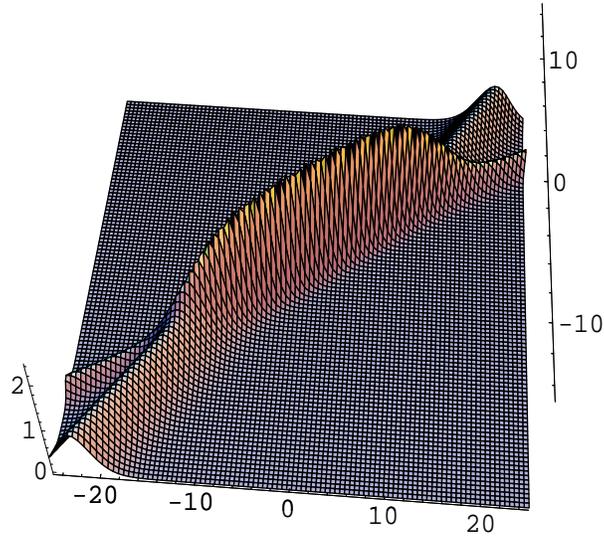,height=8cm,width=8cm}
\end{center}
\caption{3D plot of typical soliton resonant state with one soliton resonance }
\end{figure}

\begin{figure}[h]
\begin{center}
\epsfig{figure=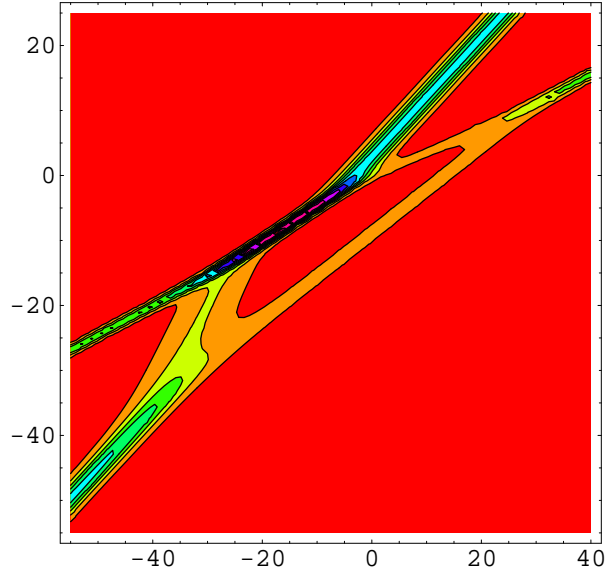,height=8cm,width=8cm}
\end{center}
\caption{Contour plot of four soliton resonances }
\end{figure}

\section{Conclusions}

The problem of chiral solitons in quantum potential, as a reduction of 2+1 dimensional Chern-Simons theory, was formulated in terms of
family of integrable derivative NLS equations by the Madelung fluid representation. By using new, non-Madelung fluid representation we
constructed integrable family of hydrodynamical systems of the Kaup-Broer type. By bilinear method we found resonance character of
corresponding chiral soliton mutual interaction.

\section{Acknowledgements}

This work was supported partially by Institute of Mathematics, Academia Sinica, Taipei, Taiwan and
Izmir Institute od Technology, Izmir, Turkey.


\end{document}